# Tunable broadband light emission from graphene

Lavinia Ghirardini[1,§], Eva A. A. Pogna[1,2,§], Giancarlo Soavi[3,4,§], Andrea Tomadin[5], Paolo Biagioni[1], Stefano Dal Conte[1], Domenico De Fazio[3], T. Taniguchi,[6] K. Watanabe,[7] Lamberto Duò[1], Marco Finazzi[1], Marco Polini[8], Andrea C. Ferrari[3], Giulio Cerullo[1*], Michele Celebrano[1*]

[1]Politecnico di Milano, Physics Department , Piazza Leonardo Da Vinci 32, 20133 Milano (Italy)

[2]CNR-NANO, NEST, Piazza San Silvestro 12, I-56127 Pisa

[3]Cambridge Graphene Centre, University of Cambridge, 9 JJ Thomson Avenue, Cambridge CB3 0FA, UK

[4]Institute for Solid State Physics, Abbe Center of Photonics, Friedrich-Schiller-University Jena, Max-Wien-Platz 1, 07743, Jena, Germany

[5]Physics Department, Pisa University, Largo Bruno Pontecorvo 3, 56127 Pisa, Italy

[6]International Center for Materials Nanoarchitectonics, National Institute for Materials Science, 1-1 Namiki, Tsukuba 305-0044, Japan

[7]Research Center for Functional Materials, National Institute for Materials Science, 1-1 Namiki, Tsukuba 305-0044, Japan

[8]Istituto Italiano di Tecnologie, Via Morego, 30 16163 Genova, Italy

[§]These authors equally contributed to this work.

[*]email: michele.celebrano@polimi.it; giulio.cerullo@polimi.it

**Abstract**

Graphene is an ideal material for integrated nonlinear optics thanks to its strong light-matter interaction and large nonlinear optical susceptibility. Graphene has been used in optical modulators, saturable absorbers, nonlinear frequency converters, and broadband light emitters. For the latter application, a key requirement is the ability to control and engineer the emission wavelength and bandwidth, as well as the electronic temperature of graphene. Here, we demonstrate that the emission wavelength of graphene's broadband hot carrier photoluminescence can be tuned by integration on photonic cavities, while thermal management can be achieved by out-of-plane heat transfer to hexagonal boron nitride. Our results pave the way to graphene-based ultrafast broadband light emitters with tunable emission.

# 1. Introduction

The opto-electronic properties of graphene [1] are ideal for a variety of applications [2], such as optical modulators [3-5], saturable absorbers [6], plasmonic devices [7,8], and various types of broadband detectors [9], working from the THz [10] to the visible(VIS)/infrared(IR) spectral range [11-13]. Despite being atomically thin, single-layer graphene (SLG) absorbs 2.3% per layer in the VIS (400-750 nm)[14] and near-IR (NIR, 1000-2400 nm)[15], and has a large nonlinear optical response (e.g. third order nonlinear susceptibility $\chi^{(3)} \sim 10^{-19} - 10^{-16}$ m$^2$V$^{-2}$ in the VIS and NIR depending on excitation photon energy and doping [16, 17]), paving the way to new applications such as gate-tunable frequency converters [16 - 18] and broadband light-emitters [19 - 27].

In intrinsic SLG, due to the absence of a gap and the linear dispersion of conduction and valence bands around the Dirac points, the massless Dirac Fermions have a spectrally flat optical response [14, 15, 28]. In principle, the gapless nature of SLG along with the ultrafast charge carrier recombination are expected to inhibit radiative recombination processes [19], while black-body radiation can still occur due to intra-band radiative transitions, Fig. 1a. Nevertheless, broadband light emission from excited charge carriers, referred to as hot carrier photoluminescence (HotPL), can be achieved in SLG via radiative recombination of hot electrons (HEs) [19,20] characterized by a non-equilibrium thermal distribution [16,17,30]. Following ultrafast-light absorption in SLG by vertical transitions within the Dirac cones, electrons are brought in a non-equilibrium thermal distribution (see Fig. 1b), which thermalizes via electron-electron scattering in a very short time (<20 fs) [31] into a hot Fermi Dirac (FD) distribution (i.e. a non-equilibrium thermal distribution) with electron temperature, $T_e$, that can reach 2000-3000 K [16, 17, 19, 31] (see Fig. 1 c). The HEs then thermalize with the lattice on a ps timescale [31,32], via emission of optical and acoustic phonons [31,32,33]. Due to the short lifetime of HEs in SLG [31, 32], thermal emission via HotPL has the advantage of allowing ultra-fast modulation, in principle up to frequencies of tens of GHz, comparable to the inverse of the HE lifetime[26].

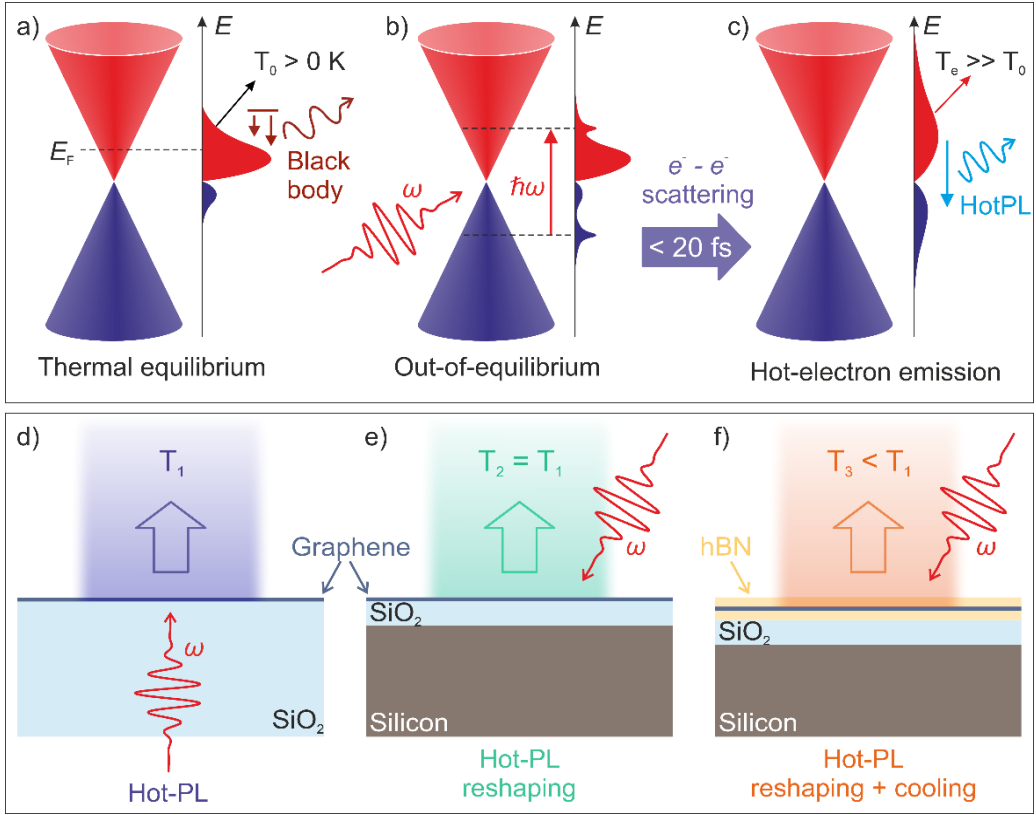

**Figure 1. HotPL emission mechanism and tuning principles.** a-c) Sketches of n-doped SLG conical bands in the vicinity of the Dirac points and of e (red)/h(blue) density distribution (i.e. the e/h distribution multiplied by the density of states), at various stages of the photoexcited HotPL dynamics. a) The electron density distribution intrinsic broadening for $T_0 > 0$ K causes black-body emission in SLG. b) Upon ultrafast laser excitation, out-of-equilibrium charge distributions are formed on sub-20-fs time scales. c) At time scales comparable with our pulse duration the e distribution thermalizes, resulting in $T_e \gg T_0$ and in two different chemical potentials for e/h. The increased h density in the valence band compared to the equilibrium situation (a) allows for the radiative recombination of hot e, leading to photoluminescence emission (HotPL). d) HotPL emission upon ultrafast laser excitation in SLG on $SiO_2$ at $T_e = T_1 \gg T_0$. e) The same process in SLG on $SiO_2$/Si substrate at the same $T_e = T_2 = T_1$, but gets spectrally reshaped by cavity effects. f) HotPL is further tuned when SLG is encapsulated in a double layer of hBN, thanks to the cooling effects between hot e and hyperbolic photons in hBN, which lead to $T_e = T_3 < T_1$.

Here, we investigate the nonlinear light emission of SLG, following NIR excitation with 150 fs pulses at 1554 nm (0.8 eV), showing emission wavelength and bandwidth tuning of the HotPL modulated by both photonic cavity effects and $T_e$ engineering, Fig. 1d-f. To this aim, we compare samples produced by micromechanical cleavage (MC) of bulk crystals [34], both pristine (MC-SLG) and encapsulated (hBN/SLG/hBN) in two ~ 10-nm-thick layers of hBN transferred on a 285-nm-thick $SiO_2$ layer on Si, with SLG grown by chemical vapor deposition (CVD-SLG), transferred either on $SiO_2$/Si or on glass. This allows us to simultaneously investigate the effect of substrate, defects and $T_e$ on the SLG HotPL. The HotPL spectral emission profile is reshaped by photonic cavity effects induced by the substrate (see Fig. 1e), as confirmed by a model describing the SLG HotPL combining Finite Difference Time Domain (FDTD) simulations, accounting for the photonic response of the substrate, with a semiclassical expression for the emission spectrum. We also demonstrate, through both HotPL and third harmonic generation (THG) measurements, that the presence of hBN allows to further tune the HotPL emission, thanks to the cooling process occurring between the hyperbolic phonons in hBN and the SLG HE, Fig. 1f. Our approach can describe the incoherent emission from any layered material (LM) coupled to any substrate and optical environment and will be thus relevant for the design of photonic devices based on LMs.

## 2. Results and discussion

### 2.1. Fabrication and characterization of the SLG devices

CVD-SLG is grown on Cu as described in Ref. [35]. A Cu foil (99.8% pure) is placed in a furnace and annealed at 1000 °C in a 20 standard cubic centimetres per minute (sccm) hydrogen flux at ~ 196 mTorr for 30mins. The growth is initiated by adding a 5 sccm $CH_4$ for 30 min. SLG is then transferred on 200-μm-thick $SiO_2$ coverslips or 285-nm-thick $SiO_2$ on Si by polymer-assisted Cu wet etching [36], using polymethyl methacrylate (PMMA). Flakes of hBN and SLG are also prepared by MC [34] of bulk graphite (NGS Naturgraphit) and hBN using adhesive tape. hBN crystals are grown under high pressure and high temperature conditions, as in Ref. [37]. SLG and hBN flakes are identified prior to transfer by a combination of optical microscopy [38] and Raman spectroscopy [39-42]. We use ~ 10-nm-thick hBN flakes as they provide sufficient screening from charge impurities in the underlying substrate [43,44]. The Raman spectra of all samples, measured with a Renishaw Raman inVia spectrometer equipped with a 50x objective at 514 nm, are reported in Fig. 2. The orange curve shows the Raman spectrum of SLG on Cu before transfer, after subtraction of the PL from Cu [45]. The spectrum shows a single sharp Lorentzian 2D peak with full width at half maximum FWHM(2D)~33 cm$^{-1}$ and peak position Pos(2D)~ 2710 cm$^{-1}$, which is a fingerprint of SLG [40]. The D peak is negligible. In hBN/SLG/hBN, the hBN E$_{2g}$ Raman peak is a combination of those from both top and bottom hBN. This yields a single peak with position Pos(E$_{2g}$)~1367cm$^{-1}$ and FWHM(E$_{2g}$)~9.8cm$^{-1}$, as expected for bulk hBN[39, 41]. Table 1 summarizes the Raman fit parameters. From these, we estimate the doping [46,47] and defect density ($n_D$) [48,49].

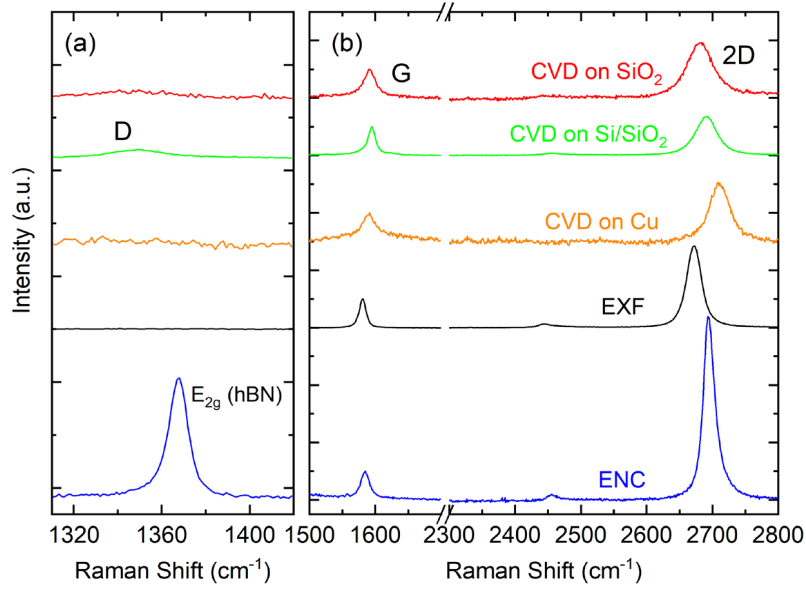

**Figure 2. Samples characterization.** Raman spectra of SLG samples at 514 nm excitation wavelength. Spectra in (a) cover the spectral region of the hBN E2g peak and of the SLG D peak, while spectra in (b) show the SLG G and 2D peaks. All spectra in (a) and (b) are normalized to the G peak intensity.

| Sample | FWHM (2D) cm$^{-1}$ | Pos (2D) cm$^{-1}$ | FWHM(G) cm$^{-1}$ | Pos(G) cm$^{-1}$ | I(2D)/I(G) | A(2D)/A(G) | E$_F$ meV | I(D)/I(G) |
|---|---|---|---|---|---|---|---|---|
| **CVD-SLG on SiO$_2$ (red)** | 45 | 2681 | 21 | 1591 | 2 | 4.4 | 250 | 0.14 |
| **CVD-SLG on Si/SiO$_2$ (green)** | 39 | 2690 | 14 | 1595 | 1.5 | 4.3 | 300 | 0.16 |
| **MC-SLG on Si/SiO$_2$ (black)** | 26 | 2672 | 12 | 1581 | 2.9 | 6.4 | 100 | |
| **hBN/SLG/hBN on Si/SiO$_2$ (blue)** | 20 | 2695 | 14 | 1584 | 7 | 9.8 | <50 | |

**Table 1.** Raman fitting parameters.

To verify the optical uniformity and spatial homogeneity of the samples, THG maps are acquired using a nonlinear optical microscope equipped with a 0.7 NA air objective for both IR illumination, with 150-fs pulses at 1554 nm from an Er- doped fiber laser, and light collection. The power impinging on the samples ranges from 300 to 700 μW (pump fluence ~ 2.1–4.9 J/m$^2$). The sample is mounted upside-down on a piezo stage to acquire nonlinear emission spatial maps. The nonlinear emission, besides that of the SLG sample deposited on glass, is collected by the objective in reflection and deflected to the detection path by a dichroic mirror (cut-off wavelength ~ 1000 nm) to reject the residual laser radiation. A flip mirror allows selecting between a VIS-NIR spectrometer, to record the whole emission spectrum, and a Si-based Single Photon Avalanche Detector (SPAD) for THG detection. For THG measurements a narrow-band filter (NBF: 520 nm (~ 2.4 eV) central wavelength – 40 nm (~ 0.18 eV) bandwidth) is in front of the SPAD. All samples are excited from the air side. THG maps are acquired by imaging 80 × 80 μm$^2$ areas on the SLG edge to enable simultaneous recording of the SiO$_2$ signal as a background to be subtracted from the total signal collected on SLG.

Figs. 3a-c show 10×10 μm$^2$ areas of the overall THG maps for MC-SLG on SiO$_2$/Si (Fig. 3a), CVD-SLG on SiO$_2$/Si (Fig. 3b), and hBN/SLG/hBN on SiO$_2$/Si (Fig. 3c). The THG and PL emission from the bare SiO$_2$/Si (darker areas in the maps) are about one order of magnitude lower than the SLG signal in the spectral window set by the filter. The THG conversion efficiency (THGE), defined as $\eta^{THG} = I_{3\omega}/I_\omega$ [16], where $I_\omega$ and $I_{3\omega}$ are the incident and THG power, respectively, is calculated by computing the total number of photons emitted at the peak fluence~ 4.9 J/m$^2$ (700 μW), about a factor two below the damage threshold of SLG (~ 10 J/m$^2$) for 150-fs pump pulses with a repetition rate~ 80MHz. This is the fluence at which the THG signal degrades after a 1min irradiation.

The total absolute THG emitted power is estimated by taking into account the overall transmission efficiency of our apparatus and by evaluating the collected fraction of the THG power. By modeling the emission from SLG as that of a dipole parallel to the SiO$_2$-air interface by FDTD[61], we estimate an overall photon collection efficiency ~ 20%, resulting from the collection NA of our objective and the presence of the substrate. This, together with the objective transmission at the THG wavelength (~80%), the optical throughput in the detection path (~ 20%), the single-photon avalanche detector quantum yield (~ 50%) and its filling factor (~ 50%), determines a drop by over two orders of magnitude in the number of detected photons with respect to the emitted ones. Thus, the total THG emitted power $I_{3\omega}$ is ~160, 130, 45 fW for MC-SLG, CVD-SLG and hBN/SLG/hBN, respectively, for an average pump power $I_\omega$ ~ 700 μW. This defines higher-bound values for $\eta^{THG}$ ~ $2.3 \times 10^{-10}$, $1.8 \times 10^{-10}$ and $6.5 \times 10^{-11}$ for the three samples, respectively, in agreement with our previous measurements [16, 17]. $\eta^{THG}$ of MC- and CVD-SLG are comparable, while for hBN/SLG/hBN $\eta^{THG}$ is ~ 3 times lower, as confirmed by the spectra recorded around the THG emission peak in Fig. 3d. In general, $\eta^{THG}$ depends on $T_e$, on the SLG Fermi energy ($E_F$), on $I_\omega$, and on the incident photon energy $\hbar\omega$ [16]. Since $I_\omega$ and $\hbar\omega$ are the same for all the samples, we attribute the different $\eta^{THG}$ to variations of $T_e$ and $E_F$ depending on the sample characteristics.

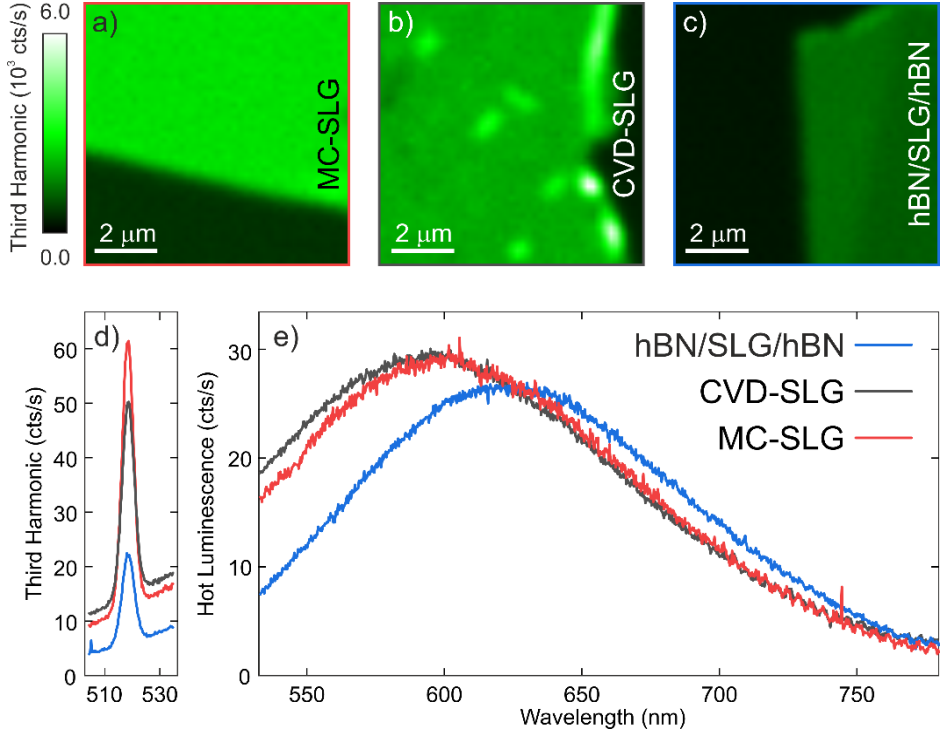

**Figure 3.** THG maps for (a) MC-SLG, (b) CVD-SLG, (c) and hBN/SLG/hBN on $SiO_2$/Si. The scanned region is $10 \times 10$ μm$^2$. d) THG peaking at ~ 518 nm for MC-SLG (red), CVD-SLG (grey) and hBN/SLG/hBN (blue). e) The visible-NIR portion of the spectrum, between 500 and 800nm, showing HotPL spectra.

## 2.2. Reshaping of the HotPL emission with a photonic cavity

To further investigate the effect of the photonic environment on the nonlinear emission properties of SLG, we also record the nonlinear emission spectra from each SLG sample with an integration time of 10s, while raster-scanning the sample over $10 \times 10$ μm$^2$. This is done to minimize photodamage and suppress the influence of intensity hot-spots resulting from nonlinear field enhancements associated with defects [62]. The nonlinear emission from the substrate, evaluated in a similar way, is then subtracted from the SLG spectra to correct for the background multi-photon PL coming from SiO$_2$ [63].

The emission spectra are shown in Figs. 3d-e and identify a THG peak~ 518 nm (~ 2.4 eV) which corresponds to one third of the 1554-nm wavelength of the pump beam, as expected for this nonlinear process [50]. The HotPL signal arising from photoexcited HEs in SLG gives a broad feature in the 500-800 nm spectral range. Refs. [19, 20, 30] reported ultrafast HotPL in MC SLG both on mica and Si/SiO$_2$ for 800 nm (1.55 eV) excitation, with spectra showing an exponential decay at increasing photon energies. A similar behavior was reported in electroluminescence experiments[24], and modeled within a semi-classical approach to account for the out-of-equilibrium condition of the system [24]. The HotPL spectra in Fig.3e show instead a maximum between 600 nm (2.06 eV) and 700 nm (1.77 eV), depending on sample and pump fluence. We attribute this behavior, which qualitatively differs from those reported in Refs. [19, 20, 25, 30], to the different excitation photon energy and sample geometry. In particular, the thickness (200-300 nm) of the SiO$_2$ on the Si substrate is comparable with the VIS and NIR wavelengths (400-1000 nm), and might lead to important interference effects, resulting in a reshaping of the HotPL spectrum [26]. Such thin-film interference effects are often exploited in photonic devices to optimize the in- and out-coupling of light [51].

To investigate the spectral reshaping of the HotPL emission from SLG, we compare the emission from two CVD SLG samples deposited on different substrates: SiO$_2$/Si or a 200 μm SiO$_2$ coverslip (see Figs. 4a,b). In the second case, the spectrum is acquired by adapting the setup in Fig. 2 to collect light in transmission. To efficiently collect the nonlinear emission, which is mostly directed towards the glass substrate, we insert a 0.85 N.A. air objective at the opposite side of the sample with respect to the illumination objective, as illustrated in Fig. 4b. The collected light is then redirected to the same detection path. The background-subtracted HotPL spectra from CVD-SLG on $SiO_2/Si$ and $SiO_2$, as a function of pump fluence, are in Figs. 4c (green) and 4d (blue), respectively.

To model the data, we first numerically estimate how the photonic environment reshapes the HotPL spectrum emitted by SLG. This is done by considering the emission from an electric dipole placed 2 nm above the substrate (either a semi-infinite glass plane or a thin $SiO_2/Si$ film). The dipole is oriented parallel to the substrate surface and its emission within the NA of the collection objective is evaluated with FDTD. This gives a set of spectral response functions, $S(\lambda)$, for the different substrates (Fig. 4e). For a semi-infinite $SiO_2$ substrate the spectral response is almost flat in the investigated wavelength range (blue line), while the emission is significantly reshaped for $SiO_2/Si$ (light green line), which acts as a broadband photonic cavity.

We apply a similar approach to evaluate the reshaping effect for the hBN/SLG/hBN. In this case the simulated emitting dipole is inserted in the center of a 20nm hBN slab with refractive index $n = 2.17$ [52,53]. The presence of hBN shifts the position of the dipole with respect to the $SiO_2/Si$ photonic cavity, resulting in a red shift of the emission peak (orange line in Fig. 4e).

Semiclassical PL theories yield the profile of the emitted light, but not the overall intensity [19], as the latter depends on the interaction between e and quantized modes of the electromagnetic spectrum. Fully quantum theories [54] require the integration of quantum-mechanical equations of motion with sub-femtosecond accuracy. This is a substantial computational effort, given that HotPL is emitted for several hundreds of fs after pulsed excitation [19,20]. Following the pump pulse, the instantaneous intensity of the HotPL decreases in time, due to electron-hole (e-h) recombination and the cooling of the HE distribution. However, the time-integrated spectrum is dominated by the signal produced by the thermalized e distribution at its hottest temperature, i.e. at the end of the pump pulse

Therefore, although a full quantum treatment of the out-of-equilibrium system would be required, here we approximate it by the signal emitted by HEs immediately after photoexcitation. We express the collected HotPL intensity as the product of $S(\lambda)$ and a factor proportional to the SLG joint density of states (JDOS) at the emitted photon wavelength for a thermalized e-h distribution at $T_e$[24]:

$$I(\lambda, T_e) = S(\lambda) \cdot \xi \frac{2hc^2}{\lambda^5} \frac{1}{\left[e^{\left(\frac{hc}{2\lambda k_B T_e}\right)}+1\right]^2} + \delta . \qquad (1)$$

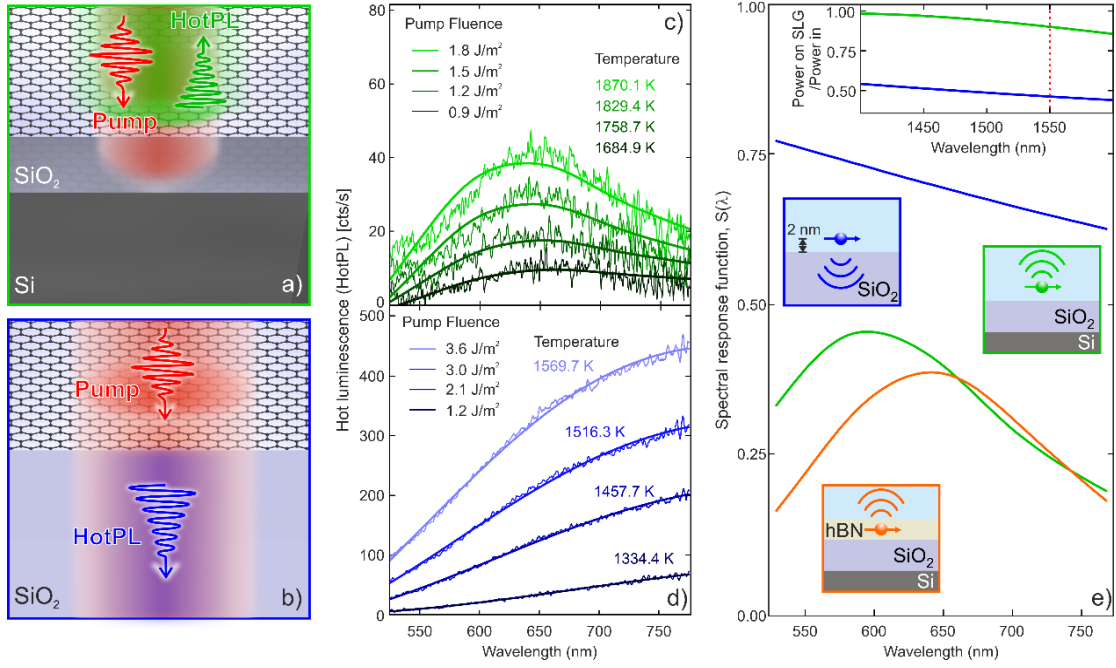

**Figure 4.** Sketches of CVD SLG on (a) SiO$_2$/Si and (b) infinitely thick SiO$_2$. Ultrafast pump pulses (red) arrive from the top using a 0.7 NA objective in both configurations, while HotPL (green and blue in a and b, respectively) is collected in back-reflection through the same objective in (a) and in transmission through a 0.85 NA objective in (b). c, d) HotPL spectra (thin lines) and fits (thick lines) from CVD-SLG for the configurations sketched in (a,b), respectively. The fits are obtained from Eq.1 by adjusting $T_e$, $\xi$ and $\delta$. $\xi$ is (c) ~ 100 and (d) ~ 4000. $\delta$ varies between -5 and -16 cts/s depending on the impinging fluence. e) $S(\lambda)$ is calculated for the various substrates using a dipole located 2 nm above SiO$_2$/Si (light green, middle inset), SiO$_2$ (blue, top inset) and inside a double layer of hBN (orange, bottom inset). Inset: pump power ratio impinging on the SLG as a function of wavelength for SiO$_2$ (blue) and SiO$_2$/Si (green), calculated by dividing the effective power deposited on SLG due to the presence of the substrate with that exiting the objective.

In Eq.1, $\xi \frac{2hc^2}{\lambda^5} \frac{1}{\left[e^{\left(\frac{hc}{2\lambda k_B T_e}\right)}+1\right]^2}$ is the JDOS multiplied by the Boltzmann distribution [24], through the dimensionless coefficient $\xi$. This depends on the oscillator strength of the direct transitions in the Dirac cones[24], parametrizing the overall intensity of the HotPL spectrum, integrated in time between pulses. Eq.1 is a reliable model for the HotPL in SLG as long as $T_e$ is the same for both e and h [16,55] and the photoexcited carrier density is much larger than the SLG doping [16,55], as in our case. In Eq. 1, $h$ is the Planck constant, $c$ the speed of light in vacuum, $k_B$ the Boltzmann constant, and $\lambda$ the wavelength. $\delta$ [cts/s] is a constant introduced to correct for residual offsets after background subtraction. For short wavelengths ($\lambda \ll \frac{hc}{2k_B T_e}$), the dominating term in Eq. 1 is an exponential tail, similar to that obtained, in such a spectral range, from black body emission [19]. The fits using Eq. 1 are superimposed to the data in Fig. 4c (solid green lines) and Fig. 4d (solid blue lines) for CVD-SLG on Si/SiO$_2$ and SiO$_2$, respectively. These are a global fit applied to each set of spectra as a function of pump fluence and treating $T_e$, $\xi$ and $\delta$ as free fitting parameters. In both CVD-SLG samples $T_e$ increases with pump fluence. Although the pump fluences for CVD-SLG on SiO$_2$/Si are lower than for CVD-SLG on SiO$_2$, the corresponding $T_e$ is higher in the first case. This can be ascribed to the influence of the photonic cavity created by SiO$_2$/Si on the 1554-nm pump beam which, as confirmed by FDTD (inset of Fig. 4e), enhances the pump fluence by a factor ~ 2. The excellent agreement between model and experiments for both data sets validates our approach.

**2.3. Tuning the HotPL emission by out-of-plane heat transfer**

To demonstrate the possibility of engineering the HotPL via out-of-plane heat transfer, we investigate with our model the HotPL emission of three different SLG samples: MC-SLG, CVD-SLG and hBN/SLG/hBN. Figs. 5a-c plot the HotPL spectra of the different samples (squares for CVD-SLG, circles for MC-SLG and triangles for hBN/SLG/hBN) measured as a function of the incident fluence. For each sample, all spectra are fitted using Eq. 1 and optimizing $T_e$, $\xi$ and $\delta$. While $\xi$ is not sensitive to fluence, $\delta$ changes with pump fluence. This could be ascribed to an imperfect background compensation. The excellent agreement between model and experiments demonstrates the robustness of this approach, once the photonic environment and the substrate are accounted for. The fluence-dependent $T_e$ plots in Fig. 5d show that MC- and CVD-SLG reach a similar $T_e$ ~ 2200-2300 K for a pump fluence~4.9 J/m$^2$, consistent with $T_e$ of Refs. [16, 19]. Despite small deviations, which we ascribe to different $E_F$, the observation of similar $T_e$ for the same illuminating fluence implies that the fabrication process and doping do not significantly affect the electron dynamics and the nonlinear optical response in our experimental conditions.

For hBN/SLG/hBN a significant drop in $T_e$ is observed (see Fig. 5d), as reflected in the red-shift of the emission spectrum (see Fig. 5c), which can only be partially attributed to the photonic response of hBN/SLG/hBN (orange curve in Fig.4c), which is different from that on bare SiO$_2$/Si (green curve in Fig. 4e). The almost 2-fold reduction in $T_e$ calls for further underlying physical processes associated with SLG encapsulation in hBN.

In hBN/SLG/hBN, out-of-plane heat transfer can occur between SLG and hBN thanks to near-field coupling between SLG HEs and the hyperbolic phonons in hBN [56,57]. This offers an efficient cooling channel, associated with a significant reduction of the SLG HE lifetime [31,56,57].

Therefore, the overall red shift in the SLG HotPL caused by the encapsulation with hBN can be interpreted as due to the combination of two effects: (i) a purely photonic one, associated with the presence of the two high refractive index hBN layers that produce an emission spectral reshaping; (ii) a more fundamental one, consisting in the modification of the thermal properties of HEs in SLG by the presence of the heat-sink consisting in the near-field coupling to hyperbolic phonons in hBN. Although a lower $T_e$ can be related to the observed HotPL red-shift (Figs. 4a-c), the quantitative evaluation of $T_e$ can be done only if the photonic response of the whole heterostructure is considered.

To further support these observations, we estimate the expected $T_e$ using the model of Refs. [58, 59] based on a set of rate equations for $T_e$ and the photoexcited carrier density in SLG, solved in the steady-state regime. Dissipation is accounted for in the relaxation-time approximation, parametrized by a cooling time $\tau$. The absorption coefficient $\alpha$ [60] includes a saturable, $T_e$-dependent contribution, due to inter-band vertical transitions, and a residual non-saturable, constant contribution $\alpha_{res}$ due to intraband transitions.

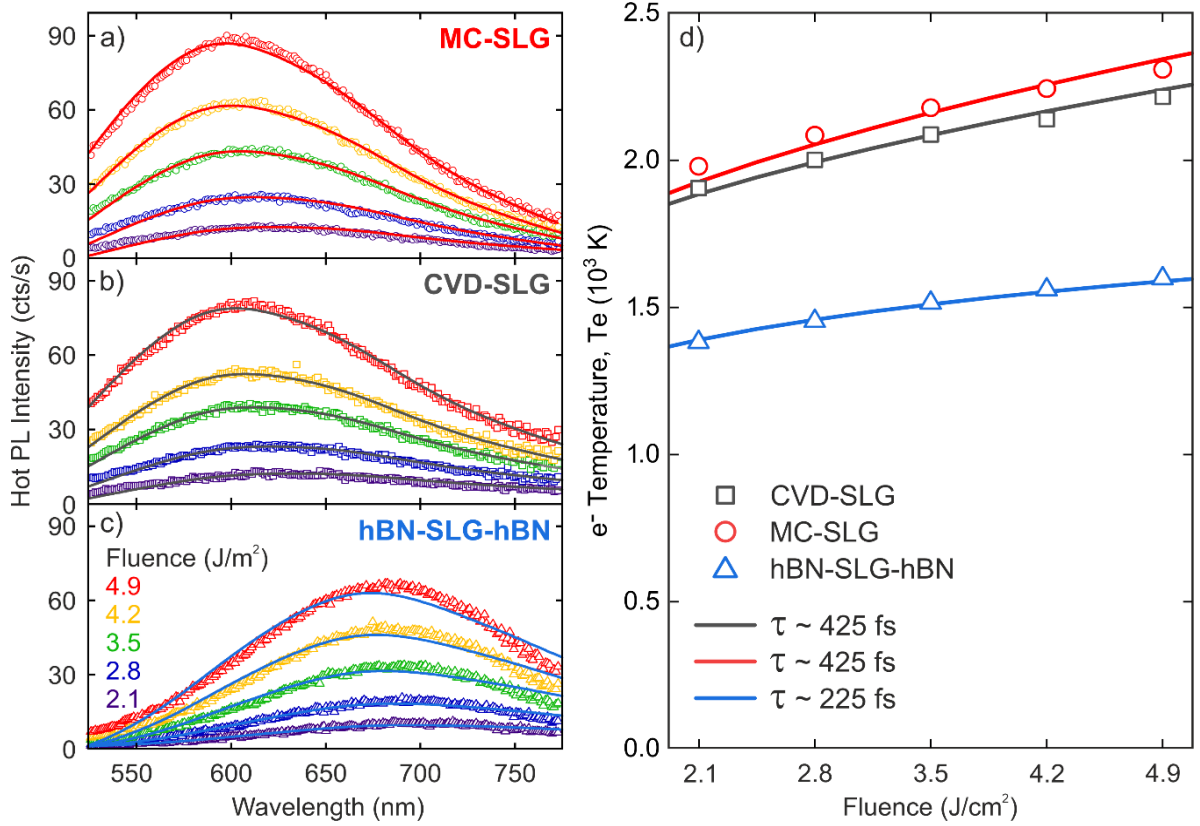

**Figure 5.** (a-c) Fluence-dependent HotPL for (a) MC-SLG, (b) CVD-SLG and (c) hBN/SLG/hBN and fits from Eq.1. The spectra in (a-c) are collected using pump fluencies ~ 2.1 (violet), 2.8 (blue), 3.5 (green), 4.2 (yellow), 4.9 J/cm$^2$ (red), the same as in (d). d) Fluence-dependence of $T_e$ extracted from the fits. (a-c) also show the fits for MC-SLG (solid red line), CVD-SLG (solid grey line) and hBN/SLG/hBN (solid light blue line), obtained varying $\delta$ from -7 to -34 cts/s. $\xi$ provides the best fits for values of 25 (MC-SLG and CVD-SLG) and 800 (hBN/SLG/hBN). The solid lines represent $T_e$ from the model in Refs. 16, 17.

We apply this model to fit the $T_e$ dependence on the illumination fluence, finding $\tau \sim 425$ fs for CVD- and MC-SLG, and $\tau \sim 225$ fs for hBN/SLG/hBN. This shorter relaxation time can be ascribed to the coupling with the hBN hyperbolic phonons. We also find $\alpha_{res} \sim 0.08\%$ for hBN/SLG/hBN, and a larger $\alpha_{res} \sim 0.4$ %, 0.5 % for CVD and MC-SLG, respectively. We use in the model a single $T_e$ for both HEs and hot h, since we assume the time-integrated spectra to be dominated by the signal produced by the thermalized e distribution at its highest $T_e$, i.e. at the end of pump pulse. Therefore, the model describes the steady state of the carriers under the action of heating, due to the pump pulse, and cooling, due to coupling with phonons.

**2.4. THG dependence on the electron temperature in SLG**

Finally, we explore the impact of the different $E_F$ and $T_e$ on THG emission, by comparing the THG fluence dependence in the three samples. After subtracting the HotPL contribution from the nonlinear emission spectra, we obtain the THG spectra in Figs.6a-c. The fluence-dependent wavelength-integrated THG powers, plotted in log-log scale in Fig. 6d, reveals a slope that exceeds the expected 3 [50], in agreement with Ref. 17, where we reported a sizable deviation from the typical cubic power law for THG caused by the dependence of the nonlinear susceptibility on both $T_e$ and $E_F$. In our experiments, $E_F \sim 0.2$ eV for CVD-SLG while $E_F \sim 0.1$ eV for MC-SLG and hBN/SLG/hBN, hence $E_F$ is much smaller than the pump photon energy (~ 0.8 eV). This condition makes the THG process extremely sensitive to $T_e$. As we discussed in Ref. 17, for $E_F < h\nu/2$, $\eta^{THG}$ is expected to grow with $T_e$, therefore resulting in a power law with an exponent larger than 3. The lower $T_e$ in hBN/SLG/hBN is accompanied by a higher THG power slope, compared to the other samples, in good agreement with the model discussed in Ref. 16. In particular, as discussed in the Supplementary Information of Ref. 16, the strongest variation of $\eta^{THG}$ upon increasing photoexcitation fluence is expected for $T_e < 1800$ K and $E_F/h\nu \ll 1$, which exactly overlap the working conditions identified for the hBN/SLG/hBN sample.

**3. Conclusions**

We measured the broadband light emission of different graphene samples following ultrashort pulse excitation. We modelled the spectra by combining the emission from an out-of-equilibrium hot electron bath in graphene and the photonic environment of graphene, also taking the response of the collection optics into account [24]. We found that, both the substrate (via photonic cavity effects) and the increase of the electronic temperature contribute to a stark shaping of the HotPL emission, as well as of the THG efficiency. We derived the dependence of the electronic temperature on the incident fluence. Under similar illumination conditions, hBN-encapsulated graphene (hBN/SLG/hBN) shows a substantially lower electronic temperature than the other samples, due to the coupling of SLG hot electrons with hBN hyperbolic phonons. The hot electron temperature influences also the dependence of the emitted third harmonic intensity on pump fluence, giving a power law with exponent>3 [17]. Thus, in SLG, both incoherent (HotPL) and coherent (THG) emission are extremely sensitive to the electronic temperature. This must be considered in ultrafast all-optical applications of SLG-based devices, such as tunable broadband light emitters and nonlinear frequency converters. Our results clarify the interplay between hot electrons and optical nonlinear effects in graphene and can be used to tailor the emission wavelength and temperature of graphene-based high-speed broadband light emitters. Our approach can be extended to reproduce the incoherent emission from any layered material coupled to any substrate.

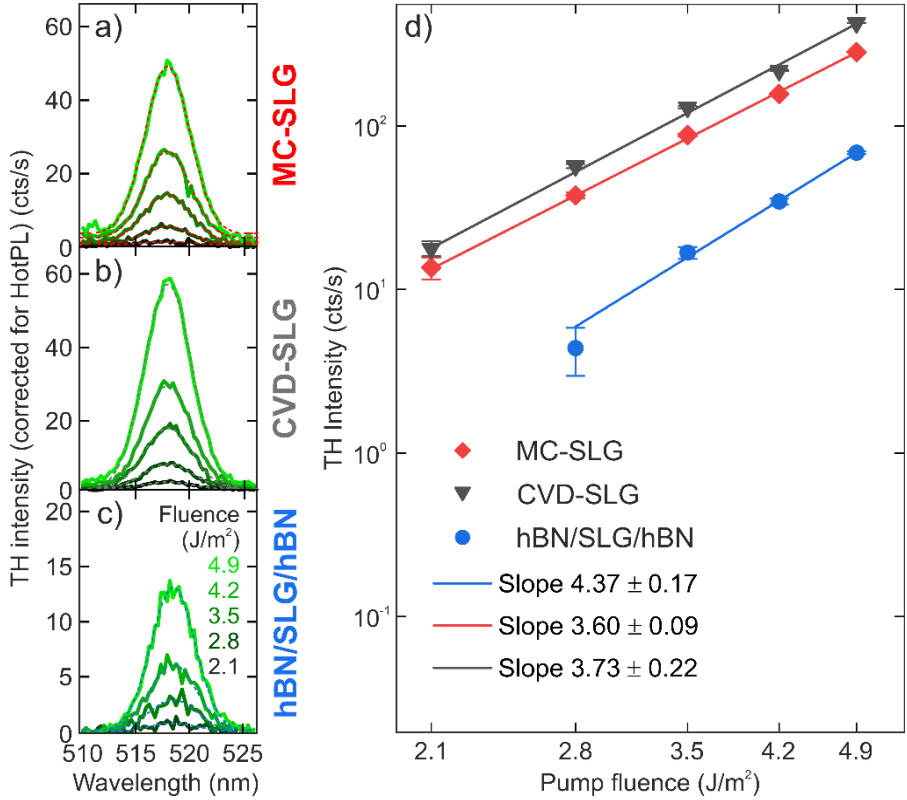

**Figure 6. Fluence dependence of TH generation.** Fluence dependent TH spectra of (a) MC-SLG, (b) CVD-SLG and (c) hBN/SLG/hBN after subtraction of HotPL background. The red, grey and blue dashed lines are Gaussian fits. (d) TH powers, measured as the areas under the Gaussian curves in panels (a-c), as a function of pump fluence. Solid lines are fits with power law curves. The error bar of the first point of hBN/SLG/hBN exits the logarithmic plot since its lower value becomes negative due to the extremely low signal levels.

**Acknowledgements**

We acknowledge funding from EU Graphene Flagship, ERC Grants Hetero2D, GSYNCOR, EPSRC Grants EP/K01711X/422/1, EP/K017144/1, EP/N010345/1, EP/L016087/1, the Elemental Strategy Initiative conducted by the MEXT, Japan, Grant Number JPMXP0112101001, JSPS KAKENHI Grant Number JP20H00354 and the CREST(JPMJCR15F3), JST.

**References**

1. Ferrari A C et al 2015 Science and technology roadmap for graphene, related two-dimensional crystals, and hybrid systems *Nanoscale* **7** 4598-4810
2. Bonaccorso F et al. 2010 Graphene photonics and optoelectronics *Nat. Photonics* **4** 611-622
3. Romagnoli M, Sorianello V, Midrio M, Koppens F H L, Huyghebaert C, Neumaier D, Galli P, Templ W, D'Errico A and Ferrari AC 2018 Graphene-based integrated photonics for next-generation datacom and telecom *Nat. Rev. Mater.* **3** 392-414.
4. Liu M et al.2011 A graphene-based broadband optical modulator *Nature* **474** 64-67
5. Sorianello V et al. 2018 Graphene-silicon phase modulators with gigahertz bandwidth *Nat. Photonics* **12** 40-44
6. Sun Z, et al. 2010 Graphene Mode-Locked Ultrafast Laser *ACS Nano* **4** 803-810
7. Grigorenko A N, Polini M, Novoselov K S 2012 Graphene plasmonics *Nat. Photonics* **6** 749–758
8. Koppens F H L, Chang D E, García de Abajo F J 2011 Graphene Plasmonics: A Platform for Strong Light–Matter Interactions *Nano Lett*. **118** 3370-3377
9. Koppens F H L, Mueller T, Avouris P, Ferrari A C F, Vitiello M S, Polini M 2014 Photodetectors based on graphene, other two-dimensional materials and hybrid systems *Nat. Nanotechnol.* **9** 780-793
10. Vicarelli L et al. 2012 Graphene field-effect transistors as room-temperature terahertz detectors *Nat. Mater.* **11** 865-871

11. Mueller T et al. 2010 Graphene photodetectors for high-speed optical communications *Nat. Photonics* **4** 297-301

12. Goykhman I et al. 2016 On-Chip Integrated, Silicon–Graphene Plasmonic Schottky Photodetector with High Responsivity and Avalanche Photogain *Nano Lett.* **16** 3005-3013

13. De Fazio D et al. 2016 High Responsivity, Large-Area Graphene/$MoS_2$ Flexible Photodetectors *ACS Nano* **10** 8252-8262

14. Nair R R, Blake P, Grigorenko A N, Novoselov K S, Booth T J, Stauber T, Peres N M R, Geim A K 2008 Fine structure constant defines visual transparency of graphene *Science* **320** 1308

15. Mak K F, Sfeir M Y, Wu Y, Lui C H, Misewich J A, Heinz T F 2008 Measurement of the Optical Conductivity of Graphene *Phys. Rev. Lett.* **101** 196405

16. Soavi G et al. 2018 Broadband, electrically tunable third-harmonic generation in graphene *Nat. Nanotechnol.* **13** 583-588

17. Soavi G et al. 2019 Hot Electrons Modulation of Third-Harmonic Generation in Graphene *ACS Photonics* **6** 2841–2849

18. Jiang T et al. 2018 Gate-tunable third-order nonlinear optical response of massless Dirac fermions in graphene *Nat. Photonics* **12** 430-436

19. Lui C H et al. 2010 Ultrafast Photoluminescence from Graphene *Phys. Rev. Lett.* **105** 127404

20. Liu W-T et al. 2010 Nonlinear broadband photoluminescence of graphene induced by femtosecond laser irradiation *Phys. Rev. B* **82** 081408(R)

21. Freitag M et al. 2010 Thermal infrared emission from biased graphene *Nat. Nanotechnol.* **5** 497-501

22. Chen C-F et al.2011 Controlling inelastic light scattering quantum pathways in graphene *Nature* **471** 617-620

23. Li T et al. 2012 Femtosecond Population Inversion and Stimulated Emission of Dense Dirac Fermions in Graphene *Phys. Rev. Lett.* **108**, 167401

24. Kim Y D et al. 2015 Bright visible light emission from graphene *Nat. Nanotechnol.* **10**, 676

25. Huang D et al. 2018 Gate Switching of Ultrafast Photoluminescence in Graphene *Nano Lett.* **18** 7985

26. Kim Y D et al. 2018 Ultrafast Graphene Light Emitters *Nano Lett.* **18** 934-940

27. Shiue R-J et al. 2019 Thermal radiation control from hot graphene electrons coupled to a photonic crystal nanocavity *Nat. Commun.* **10** 109

28. Mak K F, Ju L, Wang F, Heinz T F 2012 Optical spectroscopy of graphene: From the far infrared to the ultraviolet *Solid State Commun.* **152** 1341-1349

29. Castro Neto A H, Guinea F, Peres N M R, Novoselov K S, Geim A K 2009 The electronic properties of graphene *Rev. Mod. Phys.* **81** 109

30. Stöhr R J, Kolesov R, Pflaum J, and Wrachtrup J 2010 Fluorescence of laser-created electron-hole plasma in graphene *Phys. Rev. B* **82** 121408(R)

31. Brida D, Tomadin A, Manzoni C, Kim Y J, Lombardo A, Milana S, Nair R R, Novoselov K S, Ferrari A C, Cerullo G and Polini M 2013 Ultrafast collinear scattering and carrier multiplication in graphene *Nat. Commun.* **4** 1987

32. Breusing M, Kuehn S, Winzer T, Malić E, Milde F, Severin N, Rabe J P, Ropers C, Knorr A, and Elsaesser T 2011 Ultrafast nonequilibrium carrier dynamics in a single graphene layer *Phys. Rev. B* **83** 153410

33. Lazzeri M, Piscanec S, Mauri F, Ferrari A C, and Robertson J 2005 Electron Transport and Hot Phonons in Carbon Nanotubes *Phys. Rev. Lett.* **95** 236802

34. Novoselov K S, Jiang D, Schedin F, Booth T J, Khotkevich V V, Morozov S V, and Geim A K 2005 Two-dimensional atomic crystals *Proc. Natl. Acad. Sci. U.S.A.* **102** 10451

35. Li X, Cai W, An J, Kim S, Nah J, Yang D, Piner R, Velamakanni A, Jung I, Tutuc E, Banerjee S K, Colombo L and Ruoff R S 2009 Large-Area Synthesis of High-Quality and Uniform Graphene Films on Copper Foils *Science* **324**, 1312-1314

36. Bonaccorso F, Lombardo A, Hasan T, Sun Z, Colombo L, Ferrari A C 2012 Production and processing of graphene and 2d crystals *Mater. Today* **15** 564

37. Taniguchi T and Watanabe K 2007 Synthesis of High-Purity Boron Nitride Single Crystals under High Pressure by using Ba–BN Solvent *J. Cryst. Growth* **303** 525-529

38. Casiraghi C, Hartschuh A, Lidorikis E, Qian H, Harutyunyan H, Gokus T, Novoselov K S and Ferrari A C 2007 Rayleigh Imaging of Graphene and Graphene Layers *Nano Lett.* **7** 2711-2717

39. Reich S, Ferrari A C, Arenal R, Loiseau A, Bello I, Robertson J 2005 Resonant Raman scattering in cubic and hexagonal boron nitride *Phys. Rev. B* **71** 205201

40. Ferrari A C et al. 2006 Raman Spectrum of Graphene and Graphene Layers *Phys. Rev. Lett.* **97** 187401

41. Arenal R, Ferrari A C, Reich S, Wirtz L, Mevellec J-Y, Lefrant S, Rubio A, and Loiseau A 2006 Raman spectroscopy of single-wall boron nitride nanotubes *Nano. Lett.* **6** 1812–1816

42. Purdie D G, Pugno N M, Taniguchi T, Watanabe K, Ferrari A C and Lombardo A 2018 Cleaning interfaces in layered materials heterostructures *Nature Commun.* **9** 5387

43. Hong X, Zou K, and Zhu J 2009 Quantum scattering time and its implications on scattering sources in graphene *Phys. Rev. B* **80** 241415

44. Burson K M, Cullen W G, Adam S, Dean C R, Watanabe K, Taniguchi T, Kim P, and Fuhrer M S 2013 Direct imaging of charged impurity density in common graphene substrates *Nano Lett.* **13** 3576-3580

45. Lagatsky A A et al 2013 2 μm solid-state laser mode-locked by single-layer graphene *Appl. Phys. Lett.* **102** 013113

46. Das A, Pisana S, Chakraborty B, Piscanec S, Saha S K, Waghmare U V, Novoselov K S, Krishnamurthy H R, Geim A K, Ferrari A C, and Sood A K 2008 Monitoring dopants by Raman scattering in an electrochemically top-gated graphene transistor *Nat. Nanotechnol.* **3** 210-215

47. Basko D M, Piscanec S, and Ferrari A C 2009 Electron-electron interactions and doping dependence of the two-phonon Raman intensity in graphene *Phys. Rev. B* **80** 165413

48. Cançado L G, Jorio A, Martins Ferreira E H, Stavale F, Achete C A, Capaz R B, Moutinho M V O, Lombardo A, Kulmala T S and Ferrari A C 2011 Quantifying Defects in Graphene via Raman Spectroscopy at Different Excitation Energies *Nano Lett.* **11** 3190-3196

49. Bruna M, Ott A K, Ijäs M, Yoon D, Sassi U, Ferrari A C 2014 Doping Dependence of the Raman Spectrum of Defected Graphene *ACS Nano* **8**, 7432-7441

50. Boyd R 2008 Nonlinear Opitcs 3$^{rd}$ edition Academic Press, Cambridge, Massachusetts

51. Roddaro S, Pingue P, Piazza V, Pellegrini V, Beltram F 2007 The Optical Visibility of Graphene: Interference Colors of Ultrathin Graphite on $SiO_2$ *Nano Lett.* **79**, 2707-2710

52. Schubert M, Rheinlander B, Franke E, Neumann H, Hann J, Order M, Richter F 1997 Anisotropy of boron nitride thin-film reflectivity spectra by generalized ellipsometry *Appl. Phys. Lett.* **70** 1819

53. Rah Y, Jin Y, Kim S and Yu K 2019 Optical analysis of the refractive index and birefringence of hexagonal boron nitride from the visible to near-infrared *Opt. Lett.* **44** 3797-3800

54. Kira m and Koch S W 2012 Semiconductor Quantum Optics (Cambridge University Press, Cambridge)

55. Tomadin a et al. 2018 The ultrafast dynamics and conductivity of photoexcited graphene at different Fermi energies *Science Advances* **4**, eaar5313

56. Tielrooij K-J et al. 2018 Out-of-plane heat transfer in van der Waals stacks through electron–hyperbolic phonon coupling *Nat. Nanotechnol.* **13** 41-46

57. Baudin E, Voisin C and Placais B 2020 Hyperbolic Phonon Polariton Electroluminescence as an Electronic Cooling Pathway *Adv. Funct. Mater.* **30** 1904783

58. Rana F, George P A, Strait J H, Dawlaty J, Shivaraman S, Chandrashekhar M and Spencer M G 2009 Carrier recombination and generation rates for intravalley and intervalley phonon scattering in graphene *Phys. Rev. B* **79** 115447

59. Wang H, Strait J H, George P A, Shivaraman S, Shields V B, Chandrashekhar M, Hwang J, Rana F, Spencer M G, Ruiz-Vargas C S and Park J 2010 Ultrafast relaxation dynamics of hot optical phonons in graphene *Appl. Phys. Lett.* **96** 081917

60. Grosso G and Pastori Parravicini G 2013 Solid State Physics (Academic Press)

61. Sullivan D M 2013 Electromagnetic Simulation Using the FDTD Method 2$^{nd}$ edition (John Wiley and Sons, Inc., Hoboken, New Jersey)

62. Bozhevolnyi S I, Beermann J and Coello V 2003 Direct Observation of Localized Second-Harmonic Enhancement in Random Metal Nanostructures *Phys. Rev. Lett.* **90** 197403

63. Baraban A P et al. 2019 Luminescence of $SiO_2$ layers on silicon at various types of excitation *J. Lumin.* **205** 102-108